\documentclass[%
 aip,
 amsmath,amssymb,
 reprint,%
]{revtex4-1}

\usepackage{graphicx}
\usepackage{dcolumn}
\usepackage{bm}

\usepackage[utf8]{inputenc}
\usepackage[T1]{fontenc}
\usepackage{mathptmx}

\begin{document}

\preprint{AIP/123-QED}

\title{A temperature-modulated dilatometer by using a piezobender-based device}

\author{Yanhong Gu}
\affiliation{Beijing National Laboratory for Condensed Matter Physics, Institute of Physics, Chinese Academy of Sciences, Beijing 100190, China}
\affiliation{School of Physical Sciences, University of Chinese Academy of Sciences, Beijing 100190, China}
\author{Bo Liu}
\affiliation{Beijing National Laboratory for Condensed Matter Physics, Institute of Physics, Chinese Academy of Sciences, Beijing 100190, China}
\affiliation{School of Physical Sciences, University of Chinese Academy of Sciences, Beijing 100190, China}
\author{Wenshan Hong}
\affiliation{Beijing National Laboratory for Condensed Matter Physics, Institute of Physics, Chinese Academy of Sciences, Beijing 100190, China}
\affiliation{School of Physical Sciences, University of Chinese Academy of Sciences, Beijing 100190, China}
\author{Zhaoyu Liu}
\affiliation{Beijing National Laboratory for Condensed Matter Physics, Institute of Physics, Chinese Academy of Sciences, Beijing 100190, China}
\affiliation{Department of Physics, University of Washington, Seattle, WA 98195, USA}
\author{Wenliang Zhang}
\affiliation{Beijing National Laboratory for Condensed Matter Physics, Institute of Physics, Chinese Academy of Sciences, Beijing 100190, China}
\affiliation{Swiss Light Source, Paul Scherrer Institut, CH-5232 Villigen PSI, Switzerland}
\author{Xiaoyan Ma}
\email{mxy@iphy.ac.cn}
\affiliation{Beijing National Laboratory for Condensed Matter Physics, Institute of Physics, Chinese Academy of Sciences, Beijing 100190, China}
\affiliation{School of Physical Sciences, University of Chinese Academy of Sciences, Beijing 100190, China}
\author{Shiliang Li}
\email[Author to whom correspondence should be addressed:]{slli@iphy.ac.cn}
\affiliation{Beijing National Laboratory for Condensed Matter Physics, Institute of Physics, Chinese Academy of Sciences, Beijing 100190, China}
\affiliation{School of Physical Sciences, University of Chinese Academy of Sciences, Beijing 100190, China}
\affiliation{Songshan Lake Materials Laboratory , Dongguan, Guangdong 523808, China}

\date{\today}

\begin{abstract}
We report a new design of temperature-modulated dilatometer, which obtains the linear thermal expansion coefficient by measuring the oscillating changes of the sample's length and temperature by a piezobender and a thermocouple, respectively. Using an iron-based superconductor KFe$_2$As$_2$ as an example, we show that this device is able to measure thin samples with high resolutions at low temperatures and high magnetic fields. Despite its incapability of giving absolute values, the new dilatometer provides a high-resolution method to study many important physical properties in condensed matter physics, such as thermal and quantum phase transitions, and vortex dynamics in the superconducting state. The prototype design of this device can be further improved in many aspects to meet particular requirements.
\end{abstract}

\maketitle

\section{introduction}

The thermal expansion of a material describes the variation of its length or volume with temperature and is an fundamental thermodynamical parameter in studying the physical properties of solids \cite{BarronTHK80}. Especially, studying the change of thermal expansion across a phase transition has become one of the important techniques in condensed matter physics since it can reflect the intrinsic change of the electronic system with very high resolution \cite{ZhaoG97,BianchiA02,MotoyamaG03,HembergerJ07,HardyF13}. In the case of studying superconductors, the thermal-expansion measurement can detect phase transitions and vortex dynamics within the superconducting state \cite{ModlerR96,LortzR03,ZaumS11,MakHK13}, which shows its unique advantages compared to resistivity and magnetic-susceptibility measurements. The ratio between the thermal expansion and the specific heat gives the Gr{\"u}neisen parameter, which is a crucial thermodynamical parameter for studying quantum critical transitions \cite{ZhuL03,LorenzT08,DonathJG08,MeingastC12,TokiwaY13,SteppkeA13}. 

For low-temperature measurements in the field of condensed matter physics, the mostly used and accurate dilatometers are based on directly measuring the length change of a sample with temperature by a plate capacitor or based upon an atomic microscope piezocantilever \cite{SchmiedeshoffGM06,KuchlerR12,AbeS12,MannaRS12,InoueD14,KuchlerR16,KuchlerR17,WangL17}. The linear thermal-expansion coefficient $\alpha$ can be directly derived from $dL/LdT$, where $L$ and $T$ are the length and temperature of the sample, respectively. 
Using alternating-current (AC) heating method, the thermal expansion can also be measured by temperature-modulated dilatometers (TMD), but previous reports only show measurements above room temperature \cite{UchinoK80,JohansenTH87,SuJ98}. The advantage of the TMD is that the setup is simple. The difficulty is how to measure the alternating length change of a sample at low temperatures. 

In this work, we present a new design of TMD based on a piezobender device that can be used at low temperatures and high magnetic fields. It is based on the uniaxial pressure device as reported previously \cite{LiuZ16,GuY17}. With slightly changing the setup, we find that the device can be easily adapted to measure the linear thermal expansion. This method is similar to the temperature-modulated calorimetry (TMC) and other temperature-modulated measurements \cite{SullivanPF68,GmelinE97,IkedaMS19}. The basic principle is to periodically heat a sample so that the oscillating changes of a sample's length and temperature, i.e., $\Delta L$ and $\Delta T$, can be simultaneously determined with the assistance of a lock-in system. High resolutions can be thus achieved because of the capability of detecting small signals by the lock-in system. The linear thermal expansion $\alpha$ can be directly obtained as $\Delta L/L\Delta T$. We will use the iron-based superconductor KFe$_2$As$_2$ as an example to show the ability of our device of measuring the thermal expansion at low temperatures and high magnetic fields.

\section{Experimental setup and measurement principles}
\begin{figure}[tbp]
\includegraphics[width=\columnwidth]{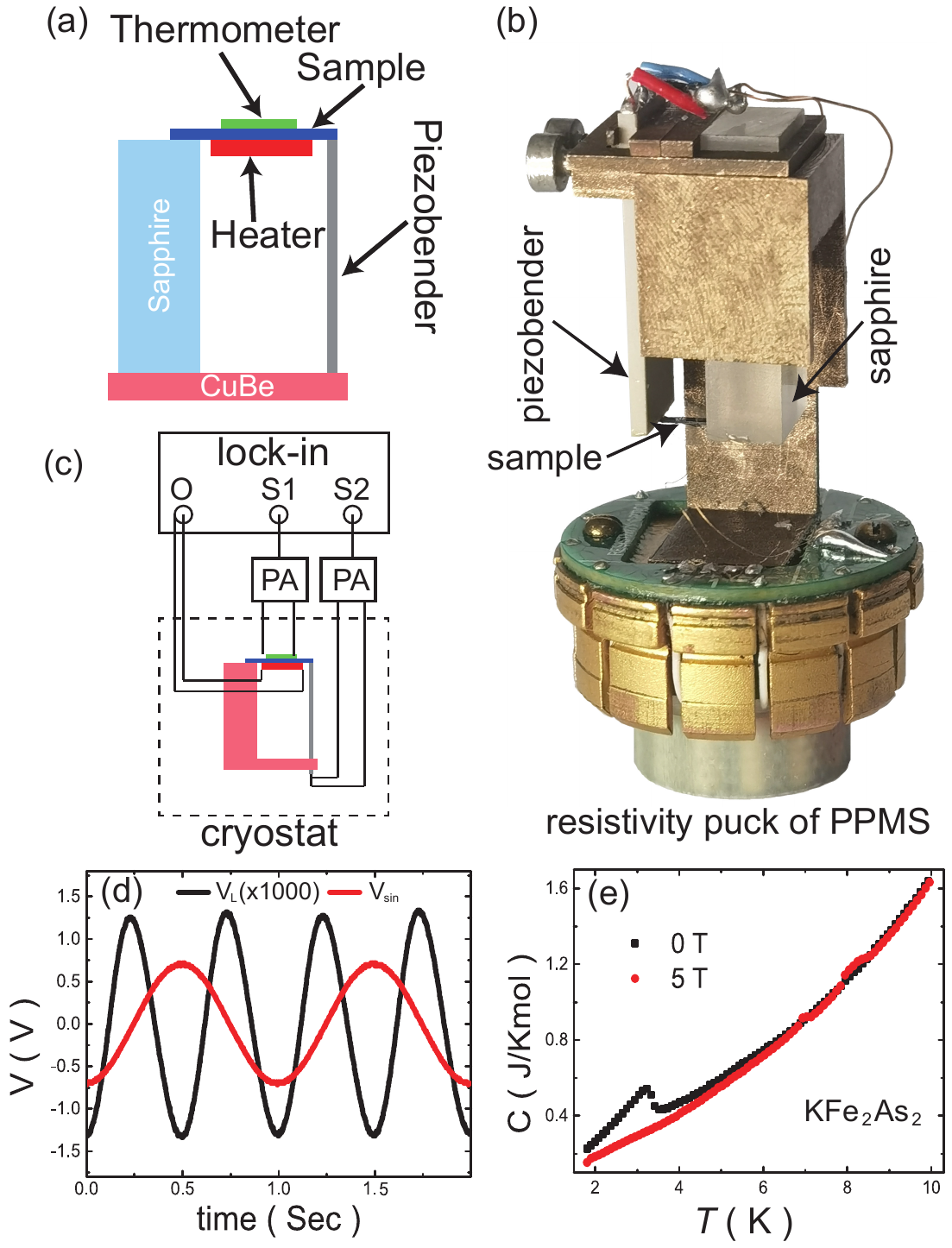}
\caption{(a) Sketch of the sample holder for AC dilatometer. The two ends of the sample are attached to the Sapphire block and piezobender, which are secured by a CuBe frame. A heater and a thermometer are attached to the two sides of the samples. (b) A photo of the AC dilatometer on the resistivity puck of the PPMS. The diameter of the puck is about 30 mm. The bottoms of the pizeobender and the sapphire are separated by several plates, whose thickness can be changed to fit the sample's length. (c) Measurement diagram of the AC dilatometer. O, S1/S2 and PA represent the sineout and input channels of the lock-in amplifier, and the pre-amplifier, respectively. (d) The comparison of the waveforms of $V_L$ (black line) and $V_{sin}$ (red line). The frequency of the sinout voltage of the lock-in amplifier is 1 Hz. The temperature is 5 K. A preamplifier with 1000 gain has been used. (e) Temperature dependence of the specific heat for KFe$_2$As$_2$ at 0 (black squares) and 5 (red circles) Tesla. }
\label{dilatometer}
\end{figure}

Figure \ref{dilatometer}(a) and \ref{dilatometer}(b) show the sketch and the photo of the AC dilatometer, respectively, which is similar to the uniaxial pressure device reported previously \cite{LiuZ16}. It is composed of a piezobender and a sapphire block secured by a CuBe frame . The two ends of a thin-slab sample are glued on top of them by GE varnish. The length of the sample is limited by the adjustable distance between the piezobender and the sapphire, which can be down to 0.5 mm for our device.  The model of the piezobender is NAC2222 (Noliac) with the length, width and thickness as 21, 7.8 and 1.3 mm, respectively. There are three contacts for the piezobender and two side ones are soldered together. In using it as a uniaxial pressure device, a DC voltage is applied to the piezobender, which will tend to move it due to the reverse piezoelectric effect and thus provide a force on the sample. To measure the thermal expansion, a heater and a thermometer are attached to two sides of the sample by N grease, as shown in Fig. \ref{dilatometer}(a). The heater can be an external heater such as a resistor heater or the sample can be heated by itself with current flowing. The thermocouple made of Ni90/Cr10 and Au99.93/Fe0.07 wires was used as the thermometer. The setup in Fig. \ref{dilatometer}(a) was put on to the regular sample puck of the Physical Properties Measurement System (PPMS, Quantum Design) as shown in Fig. \ref{dilatometer}(b), which provides the low-temperature and magnetic-field environment. A lock-in system was used to provide the AC power and measure the signals from the thermometer and piezobender. The measurement diagram is shown in Fig. \ref{dilatometer}(c). In this work, we have used a two-channel lock-in system (Model OE1022D from SYSU Scientific Instruments) and two preamplfiers (Model SR560 from Stanford Research Systems). 

Assuming that the AC current flowing though the heater with the resistance $R$ has the form of $I_{AC}sin\omega t$ and neglecting the phase change, the AC power supplied by it will be $P_{AC}sin^2\omega t$, where $P_{AC} = I_{AC}^2R$/2. This will result in an oscillating temperature of the sample, $\Delta T$ with the $2\omega$ periodicity. With proper conditions, it has already been shown that the magnitude of $\Delta T$, i.e. $\Delta T_{AC}$, has the following form \cite{SullivanPF68},
\begin{equation}
\Delta T_{AC} = \frac{P_{AC}}{2\omega C_s}\left[1+\frac{1}{4\omega^2\tau_1^2}+4\omega^2\tau_2^2+\frac{2K_b}{3K_s}\right]^{-\frac{1}{2}},
\label{fDep}
\end{equation}
\noindent where $C_s$ is the heat capacity of the sample, $K_b$ is the thermal conductance between the sample and the bath, $K_s$ is the thermal conductance of the sample, $\tau_1$ is the relaxation time from sample to bath, $\tau_2$ is associated with the time which the sample, the heater and the thermometer attain thermal equilibrium. The frequency-independent term $2K_b/3K_s$ is the geometric correction due to the finite thermal diffusivity of the sample. In the case where $\tau_1$ is large while $\tau_2$ and $2K_b/3K_s$ are small, $\Delta T_{AC}$ can take the much simpler form as 
\begin{equation}
\Delta T_{AC} = \frac{P_{AC}}{2\omega C_s},
\label{optimal}
\end{equation}
\noindent which is the case for most traditional AC calorimeters. 

The oscillating temperature of the sample will simultaneously cause the oscillation of its volume. The change of the length along a particular direction can be measured by the piezobender for the setup in Fig. \ref{dilatometer}(a) since the movement of the top will create a voltage that can be detected by the lock-in system. Figure \ref{dilatometer}(d) shows an example, where $V_{sin}$ is the sine output voltage of the lock-in amplifier while $V_{PB}$ is the voltage of the piezobender. It is clear that the periodicity of $V_{PB}$ is twice of that of $V_{sin}$. Neglecting all the phase differences, the linear expansion coefficient $\alpha$ is thus 
\begin{equation}
\alpha =\frac{\Delta L_{AC}}{L_0\Delta T_{AC}}=\frac{\eta\kappa\Delta V_{L}}{L_0\Delta T_{AC}},
\label{EqACdilatometer}
\end{equation}
\noindent where $L_0$ is the static length of the sample between the tops of the BeCu frame and the piezobender, which is assumed to be a constant value since its change with temperature can be neglected. The magnitude of the oscillating length, $\Delta L_{AC}$, is equal to $\eta\kappa\Delta V_L$, where $\Delta V_L$ is the magnitude of the oscillating voltage on the piezobender. The coefficient $\kappa$ is the relationship between the moving distance of the top of the piezobender and the voltage resulted from, which has been independently determined to be about (43.5 + 0.48T) nm/V below 10 K with the DC voltage supply \cite{mxy}. The coefficient $\eta$ is introduced to account for other factors that may affect the determination of the length change. The value of $\eta$ is hard to determine and we will give a rough estimation by comparing our results with those measured in the capacitive dilatometer.

The sample used here is an iron-based superconductor KFe$_2$As$_2$. At $T_c$ = 3.4 K, its thermal expansion along the a axis direction shows a drastic jump, changing from positive to negative value \cite{HardyF13}. We grew the KFe$_2$As$_2$ samples by the flux method as reported previously \cite{LiuZ19}. The specific heat shows a clear superconducting transition at 0 T, which disappears at 5 T, as shown in Fig. \ref{dilatometer}(e). Most of the thermal-expansion results were measured on a slice of the KFe$_2$As$_2$ sample, which was cut along the a axis with the length, width and thickness as 4, 0.85 and 0.14 mm, respectively. Two electronic contacts were made on the sample by silver epoxy. Because the resistance of the sample is very small ($<$ 10$^{-3} \Omega$), it is the contact resistances that actually work as the heater.  Some other KFe$_2$As$_2$ samples have also been measured to study the effect of sample size for some parameters.

\section{Results}

\begin{figure}[tbp]
\includegraphics[width=\columnwidth]{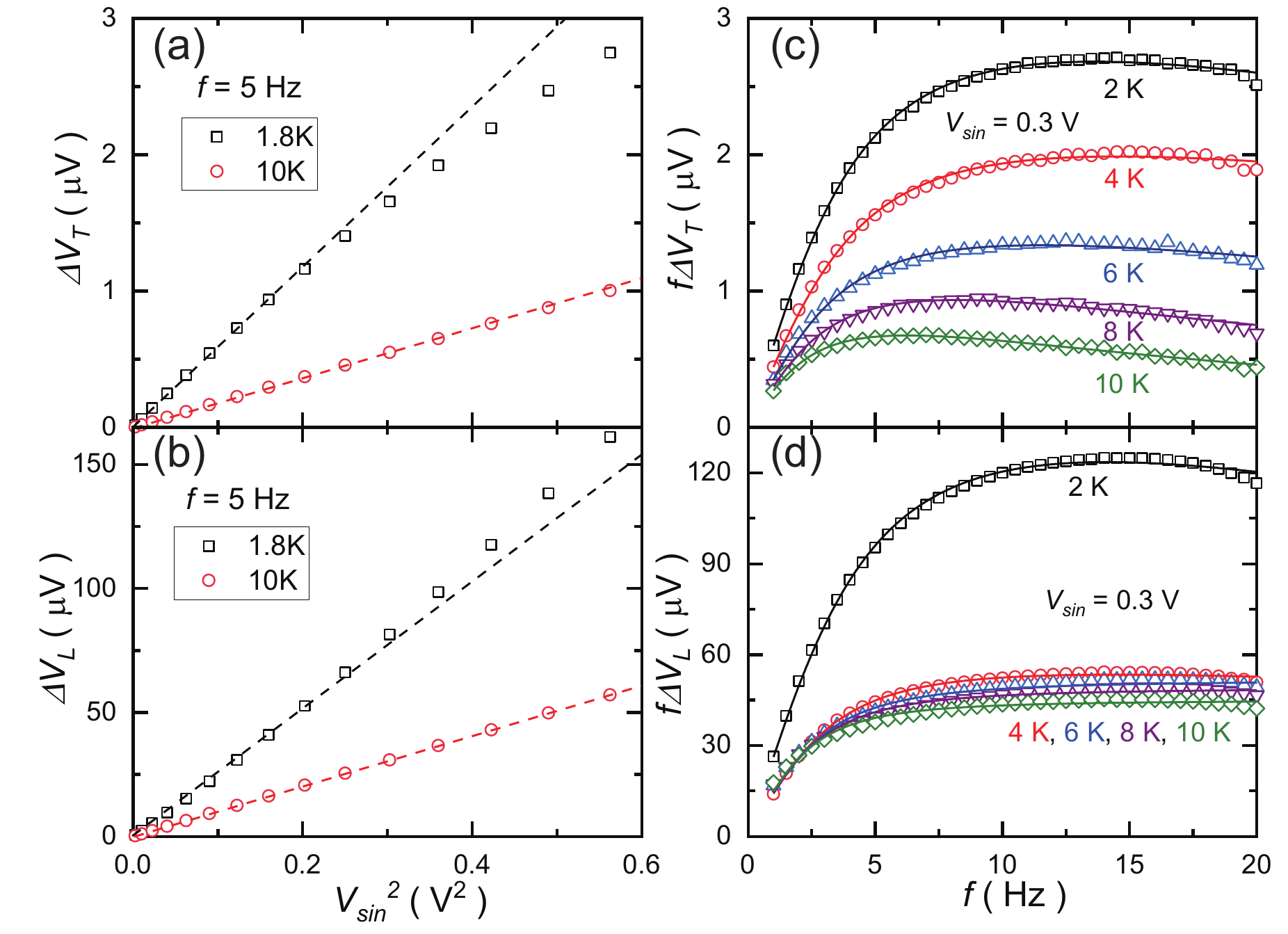}
\caption{(a) \& (b) The $V^2_{sin}$ dependence of $\Delta V_T$ and $\Delta V_L$, respectively. The frequency is fixed at 5 Hz. The temperatures are 10 and 1.8 K. The dashed lines are fitted by a linear function for $V_{sin} <$ 0.3 V. (c) and (d) The frequency dependence of $fV_T$ and $fV_L$, respectively. The solid lines are the fitted results according to Eq. \ref{fDep}.}
\label{basic}
\end{figure}

According to Eq. (2) and (3), both the magnitudes of voltages from the thermocouple and the piezobender, $\Delta V_T$ and $\Delta V_L$, should be proportional to the square of the magnitude of the heating voltage $V_{sin}$. As shown in Fig. \ref{basic}(a) and \ref{basic}(b), the quadratic relation holds for small voltages. At large voltages, $\Delta V_T$ becomes lower than the value from linear extrapolation. This is most likely due to the DC heating effect \cite{GmelinE97}, which results in an increase of the sample temperature and thus the increase of the specific heat and decrease of $\Delta T$. To avoid this effect, $V_{sin}$ is fixed as 0.3 V in the following measurements.

The frequency dependence of the $\Delta V_L$ and $\Delta V_T$ can be well described by Eq. \ref{fDep} with the term 2$K_b$/3$K_s$ neglected, as shown in Fig. \ref{basic}(c) and \ref{basic}(d), respectively. At low temperatures (2 and 4 K), $f\Delta V_T$ becomes frequency-independent at high frequencies, which is because $\tau_2$ is very small ($\sim$ 2.1 ms). With increasing temperatures, $\tau_2$ becomes larger so that its effect moves to lower frequencies and there is no frequency-independent region. For $f\Delta V_L$, the frequency-independent region exists at all temperatures. The difference of $\tau_2$ for $\Delta V_L$ and $\Delta V$ suggests that it takes longer time for the thermal couple to become thermal equilibrium since N-grease has been used to attach it to the sample. The frequency-independent region means that the signal has the form in Eq. \ref{optimal}, which suggests that the device works in the optimal condition for heat-capacity measurements in these frequencies. 

\begin{figure}[tbp]
\includegraphics[width=\columnwidth]{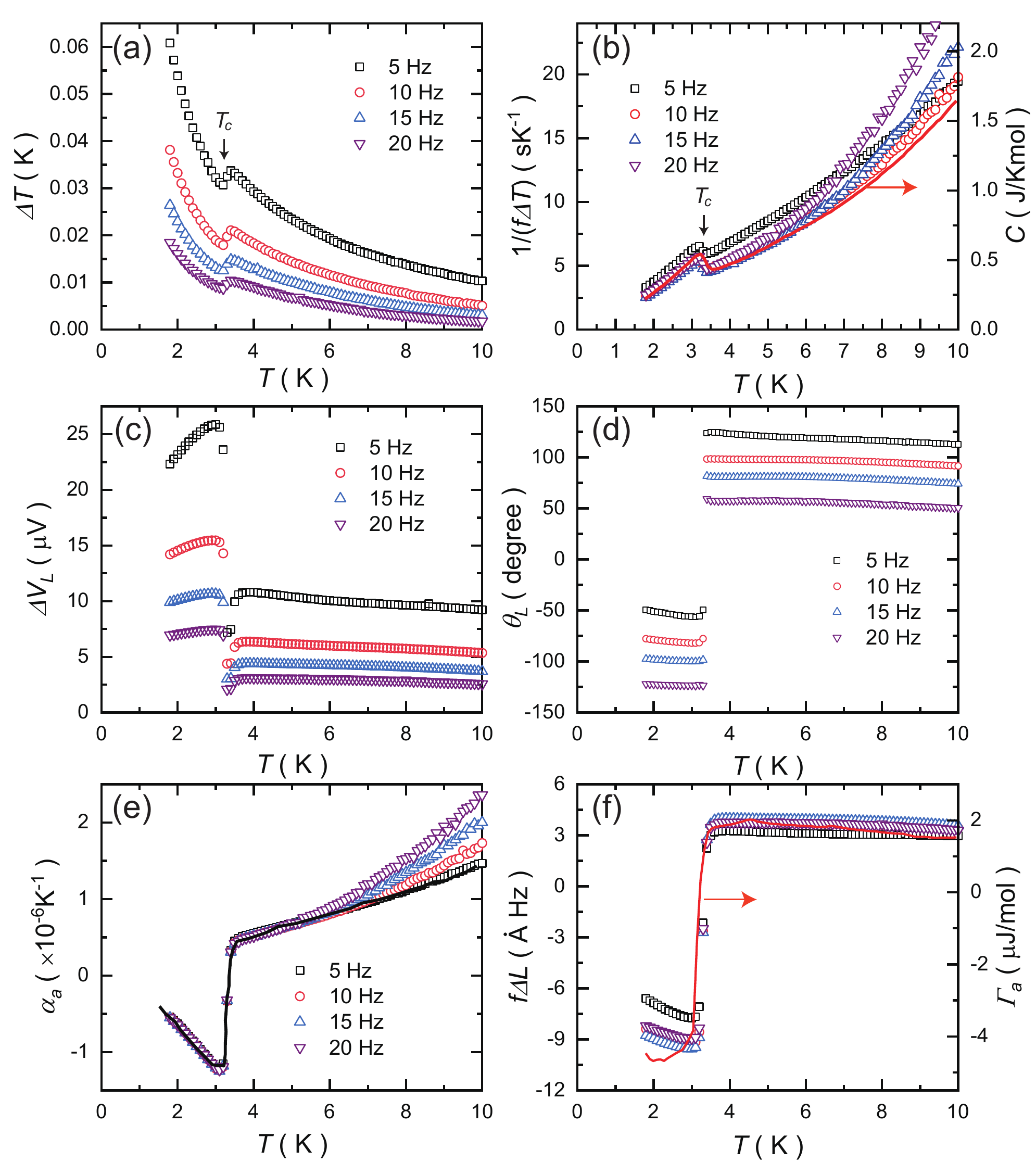}
\caption{ (a) Temperature dependence of $\Delta T$. (b) Temperature dependence of $1/(f\Delta T)$ and the specific heat (solid line). (c) \& (d) Temperature dependence of $\Delta V_L$ and $\theta_L$, respectively. (e) The temperature dependence of the linear temperature thermal expansion along the a axis $\alpha_a$. The solid line is $\alpha_a$ from Ref. \cite{HardyF13}. (f) The temperature dependence of the linear Grüneisen parameter. The solid line in (d) is $\Gamma_a$ calculated by $\alpha$ from Ref. \cite{HardyF13} and the specific heat of KFe$_2$As$_2$. }
\label{alpha}
\end{figure}

Figure \ref{alpha}(a) shows the temperature dependence of $\Delta T$, which is calculated by $\Delta V_T/S$ with $S$ as the Seebeck coefficient of the thermocouple. It increases with decreasing temperature and shows a dip at $T_c$. According to Eq. \ref{optimal}, the specific heat should be proportional to $1/(f\Delta T)$ if the power $P_{AC}$ has little temperature dependence. Figure \ref{alpha}(b) shows the comparison between $1/(f\Delta T)$ and the specific heat $C$. Below above 5 K, the data above 10 Hz closely follow the temperature dependence of $C$, which indicates the device is working at the optimal condition for the specific heat measurements. On the other hand, significant deviation occurs at higher temperatures, which is due to the increase of $\tau_2$ for $\Delta V_T$ as discussed above. 

Figure \ref{alpha}(c) shows the temperature dependence of $\Delta V_L$, which shows a very sharp dip at $T_c$ for all frequencies. This is because 
the linear thermal expansion coefficient $\alpha$ along the a axis changes sign at $T_c$ \cite{HardyF13}. As shown in Fig. \ref{alpha}(d), the phase $\theta_L$ for $\Delta V_L$ changes exactly 180 degrees at $T_c$, which demonstrates the sign change of $\Delta L$ across $T_c$. Except this 180-degree change, the value of $\theta_L$ has no physical meaning, so the change of length is only associated with $\Delta V_L$. 

Figure \ref{alpha}(e) shows the linear thermal expansion $\alpha_a$ along the a axis obtained by Eq. (3) with $\eta$ = 134. The value of $\eta$ is obtained by comparing our results with previous results measured by the capacitive dilatometer \cite{HardyF13}. The origin of the large value of $\eta$ will be discussed later. The temperature dependence of $\alpha_a$ below 5 K shows almost no frequency dependence and can be nicely normalized to the reported values. With increasing temperature, $\alpha_a$ becomes different for different frequencies, which is because the frequency dependence of $\Delta V_L$ and $\Delta V_T$ becomes different as shown in Figs. \ref{basic}(c) and \ref{basic}(d). 

Figure \ref{alpha}(f) shows the temperature dependence $f\Delta L$ = $f\eta\kappa \Delta V_L$ at different frequencies. The reason to plot this is that this value should be proportional to the linear Gr{\"u}neisen parameter $\Gamma_a$ = $\alpha_a/C_s$ according to Eq. \ref{fDep} and \ref{EqACdilatometer}.It is clear that the calculated values nicely follow the temperature dependence of $\Gamma_a$ calculated from the $\alpha_a$ in Ref. \cite{HardyF13} and the specific heat. It should be noted that in this method, there is actually no need to measure $\Delta T$. The absolute value of $\Gamma_a$ can be obtained if the power $P_{AC}$ can be determined. 

\section{Discussions}

The above results demonstrate that it is possible to measure the thermal expansion at low temperatures by the AC method based on a piezobender device. There are several advantages compared to the capacitive dilatometer. First, the design and fabrication of the sample holder, and the experimental setup are very simple. The capacitive low-temperature dilatometer on the other hand is very specialized and its measurement needs high-resolution capacitance meter. Second, the device here can measure very thin crystals, which may be crucial for some materials where only thin slices of crystals are available. With the known spring constant of the piezobender ( $\sim$ 0.0308 N/$\mu$m) and the Young's modulus ( $\sim$ 105 GPa along the a axis for KFe$_2$As$_2$ \cite{TaftiFF14}), we estimate that $F/F'$ is about 100 for our sample used here, where $F$ and $F'$ are the force created by the thermal expansion of the sample and the force required to move the tip of the piezobender, respectively. This means that although the sample is thin, its thermal expansion is able to push the piezeobender and create the voltage to be measured. In fact, the thinnest sample we have tried is about 18 $\mu$m, which still has $F/F' \approx$ 30. Third, the resolution is good as shown in Fig. \ref{alpha}(e) and (f).  Based on the relative noise level of $\Delta V_L$ ($\sim$ 10$^{-4}$) and the value of $\alpha_a$, we estimate that the absolute resolution to resolve the length change of the sample is about 10$^{-4}$ \AA  at 2 K. The resolution for $\alpha$ in this work is about 0.5 $\times$ 10$^{-9}$K$^{-1}$ at 2 K. It should be pointed out that better resolution can be achieved in the future by introducing better shielding and design of the electrical circuits, and different choices of the heater, the piezobender and the thermometer. Ideally, one may finally achieve a resolution better than 10$^{-4}$/$\sqrt{f}$ \AA at 2 K for the length change if the noise level of a lock-in system is about 2 nV/$\sqrt{f}$. 

As shown by this work, it is also possible to obtain the specific heat by our device as shown in Fig. \ref{alpha}(b). However, whether this is possible seems to depend on the sample thickness (and thus the mass), as for much thinner samples ( $\sim$ 10 $\mu$m ), the measured $C$ is about 10 times larger than the actual value, which may be because a large amount of heating power is applied on other part of the device. Nevertheless, the oscillating temperature of the sample can still be represented by $\Delta T$ and so the $\alpha$ value is still reliable. The simultaneous measurements of the linear thermal expansion $\alpha$ and the specific heat $C_s$ of a sample means that one can obtain the linear Gr{\"u}neisen parameter $\Gamma \propto \alpha/C_s$. In principle, even without the knowledge of $\Delta T$ and so the specific heat, it is still possible to directly obtain $\Gamma \propto \Delta V_L$ according to Eq. \ref{optimal} and \ref{EqACdilatometer}, as shown in Fig. \ref{alpha}(f), although it is still required that the measured $C$ is close to $C_s$. We'd like to point it out though that in some cases where both $C_s$ and $\Gamma$ diverges, such as at a quantum critical point, $\Delta V_L$ still provides a good approximation to $\Gamma$.  

There are several disadvantages for the TMD device here. First, it is hard to obtain the absolute value of $\alpha$. As show above, the values measured here are more than two orders smaller than the actual ones. We have found that the relaxation time of the charges in the piezobender, which is about 20 ms at 2 K, makes the signal 4 times smaller than the maximum voltage created. Moreover, the output voltage of a piezobender strongly depends on frequency $f$ and becomes much smaller when $f$ is away from the resonance frequency \cite{RoundyS04,AjitsariaJ07,ReillyEK11}, which is about 1300 Hz for the piezobender used here. Fortunately, our results in Fig. \ref{alpha}(f) shows that for a fixed setup, $\eta$ is independent of sample size, which means that the absolute value of $\alpha$ can be obtained by carefully comparing the results between the TMD and capacitive dilatometers. In fact, for the several KFe$_2$As$_2$ samples we have measured, the values of $\eta$ are all around 140. Second, the device works well below 5 K where the optimal working condition described by Eq. \ref{optimal} is satisfied. At higher temperatures, the effect of $\tau_2$ in $\Delta T$ becomes significant, which results in the deviation of $\alpha_a$ from the linear temperature dependence (Fig. \ref{alpha}(e)). It is interesting to note that for $\Delta V_L$, there is always a region that satisfy the optimal working condition as shown in Fig. \ref{basic}(d). It follows that the $\tau_2$ of the sample, i.e., the time which the sample attains thermal equilibrium, is much smaller than that of the thermocouple. If a thermometer with smaller heat capacity and better thermal conductance can be used, this issue may be solved. 

Despite the above disadvantages, the TMD introduced here can be used to study some important physical properties, such as the phase transitions and vortex dynamics in the superconducting state. Especially, since the device works well below 5 K, it may be possible to apply it below 1.8 K to study quantum phase transitions. Moreover, although it is hard to obtain the absolute value, studying the systematic change of the thermal expansion in a particular system with doping and magnetic field etc is still reliable. 

\section{summary}
We have designed a temperature-modulated dilatometer based on a piezobender device, which can measure the linear thermal expansion coefficient and in principle the Gr{\"u}neisen parameter. Although it is hard to obtain the absolute values, the device has the capability of measuring very thin samples with high resolutions as illustrated by measuring KFe$_2$As$_2$ single crystals. Considering that this device is still a prototype, much improvements should be possible in the future.

\begin{acknowledgments}

We thank great helps from Prof. Jing Guo. This work was supported by the National Key R\&D Program of China (Grants No. 2017YFA0302903 and No. 2016YFA0300502), the National Natural Science Foundation of China (Grants No. 11874401 and No. 11674406), the Strategic Priority Research Program(B) of the Chinese Academy of Sciences (Grants No. XDB25000000 and No. XDB07020000, No. XDB28000000). 

\end{acknowledgments}

\section{Data Availability Statement}
The data that support the findings of this study are available from the corresponding authors upon reasonable request.
%

\end{document}